\documentclass[twocolumn,showpacs,preprintnumbers,amsmath,amssymb]{revtex4}
\usepackage{amssymb,amsmath,amscd,epsfig,bbold,stmaryrd,bm,braket}
\usepackage[normalem]{ulem} % either use this (simple) or

\newcommand{\be}{\begin{equation}}
\newcommand{\ee}{\end{equation}}

\newcommand{\e}{\mathbf{e}}
\begin{document}

\title{Resonant effects in a SQUID qubit subjected to non adiabatic changes}

\author{F. Chiarello$^{1}$, S. Spilla$^{2,3}$, M. G. Castellano$^{1}$, C. Cosmelli$^{4}$, A. Messina$^{2}$, R. Migliore$^{5}$, A. Napoli$^{2}$, G. Torrioli$^{1}$}

\address{$^1$ IFN-CNR, via Cineto Romano 42, 00156 Rome, Italy}
\address{$^2$ Dipartimento di Fisica e Chimica, Universit\`a di Palermo, I-90123 Palermo, Italy}
\address{$^3$ Institut f\"ur Theorie der Statistischen Physik, RWTH Aachen University, D-52056 Aachen, Germany}
\address{$^4$ Dip. Physics, Universit\`a di Roma ``Sapienza'', 00185 Rome, Italy}
\address{$^5$ Institute of Biophysics, National Research Council via Ugo La Malfa 153, 90146 Palermo,Italy}

\date{.....}

\begin{abstract}

By quickly modifying the shape of the effective potential of a double SQUID flux qubit from a single-well to a double-well condition, we experimentally observe an anomalous behavior, namely an  alternance of resonance peaks, in the probability to find the qubit in a given flux state. The occurrence of Landau-Zener transitions as well as resonant tunneling between degenerate levels in the two wells may be invoked to partially justify the experimental results. A quantum simulation  of the time evolution of the system indeed suggests that the observed anomalous behavior can be imputable to quantum coherence effects. The interplay among all these mechanisms has a practical implication for quantum computing purposes, giving a direct measurement of the limits on the sweeping rates possible for a correct manipulation of the qubit state by means of fast flux pulses, avoiding transitions to non-computational states.\end{abstract}

\date{\today}

\include{Simboli}

\maketitle

\section{Introduction}
Superconducting devices based on the Josephson effect are an important testbed for investigating deep aspects of quantum mechanics such as macroscopic quantum phenomena \cite{MQC1}-\cite{MQC4} and circuit quantum electrodynamics (cQED) \cite{cQED1}-\cite{cQED5}. Such devices are moreover promising candidates for the practical implementation of solid state quantum computing \cite{qubit1}-\cite{qubit10}. This thanks to the possibility to arrange superconducting circuits in a desired way with great flexibility  \cite{potential1}-\cite{potential3} and also because  their behavior can be  analyzed by means of  equivalent mechanical models \cite{barone}, describing the motion of fictitious particles moving in an effective potential, with a supposed quantum behavior at low temperature.  For example, it is possible to realize Josephson anharmonic oscillators that can be used as artificial atoms, which can be manipulated by microwaves with NMR-like techniques \cite{martinis}, and which can also be coupled with superconducting resonators for single photon experiments in cavities \cite{cQED1}, \cite{cQED2}. Moreover, it is often possible to control and modify the effective potential shape in time with a fast and accurate timing. This allows, for example, the observation of very fast coherent oscillations (up to 20 GHz) of the magnetic flux states in a SQUID  (Superconducting Quantum Interference Device) qubit \cite{fastosc1}, \cite{fastosc2}, obtained just by quickly and strongly modifying the  effective  potential shape (from a symmetric double-well to a single-well, and back to the double-well). This is done by simply applying flux pulses, in the absence of microwaves.
For this kind of manipulation it is of great importance the rapidity of the modification of the  effective  potential shape. For example, if we consider quantum computing applications, the manipulation must be fast enough in order to induce non adiabatic Landau-Zener transitions between the first two energy levels (used as computational space), but also slow enough in order to avoid transitions to upper levels (non computational space). Fortunately generally speaking this is possible thanks to an appropriate energy gap existing between the first couple of levels and the upper ones, but the transition rate is an aspect that must be accurately considered and calibrated \cite{rate}.

In this work we investigate experimentally and theoretically the effect of the speed of modification of the potential shape in a double SQUID flux qubit, presenting the experimental observation and the theoretical analysis of an interesting quantum effect due to the interplay of Landau-Zener transitions and resonant tunnelling. A similar phenomenon has been studied in ref.\cite{silvestrini1}, \cite{silvestrini2}, even if in a different system and contest.

\section{The double SQUID}
\begin{figure}
\begin{center}
\includegraphics[width=8cm]{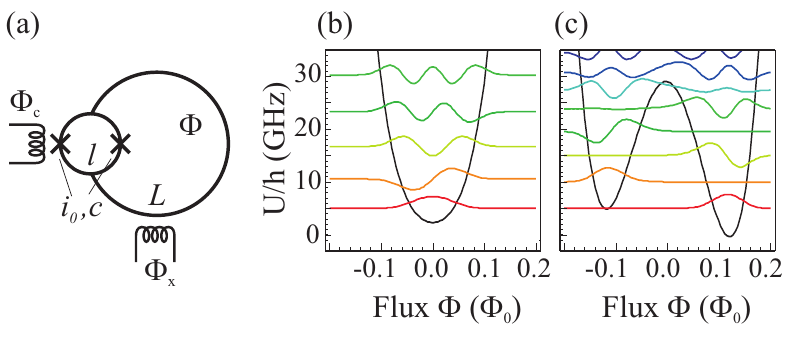}
\caption{(a) Scheme of the double SQUID. (b) Effective potential of the double SQUID in the single-well case, with relative eigenwaves vertically shifted by the corresponding eigenenergies. (c) Double-well case with a slight asymmetry.}  \label{figschema}
\end{center}
\end{figure}

The device we consider is the so called double SQUID \cite{double}, consisting of a superconducting loop of inductance $L$ interrupted by a dc SQUID, a second smaller superconducting loop of inductance $l$ interrupted by two identical Josephson junctions, each of (nominally) identical critical current $i_0$ and capacitance $c$ (fig. \ref{figschema}a). The dc SQUID behaves approximately like a single junction of capacitance $C=2c$ and tunable critical current $I_0 ( \Phi_c )=2i_0 \cos ( \Phi_c / \Phi_B )$ (where $\Phi_B=\frac{\Phi_0}{ 2 \pi}$ being $\Phi_0=h/2e$ the flux quantum), which is controlled by a magnetic flux $\Phi_c$ applied to the small loop (this approximation holds if the loop is small enough, i.e for $l i_0 \ll \Phi_0$). Note that $I_0 ( \Phi_c )$ can also be negative, and in this case the dc SQUID behaves as a pi-junction.
The double SQUID behavior can be controlled by two distinct magnetic fluxes, one applied to the large loop ($\Phi_x$) and the second to the small one ($\Phi_c$, previous mentioned). The SQUID dynamics can be described by an equivalent mechanical model, with effective mass $m=C \Phi_B^2$, effective position corresponding to the total magnetic flux threading the large loop ($\Phi$), and potential

\begin{equation}\label{H}
U = {{\left( \Phi - \Phi_x \right)^2}\over{2 L}}- {{I_0(\Phi_c) \Phi_B}} \cos \left( {\Phi \over \Phi_B }\right)
\end{equation}

This effective potential can have one or two distinct wells, according to the adimensional parameter $\beta(\Phi_c)=2 \pi I_0(\Phi_c) L/\Phi_0$: in the particular case $\Phi_x=0$ there will be a single-well for $\beta>-1$ (approximately an harmonic potential with a characteristic frequency controlled by $\Phi_c$, fig. \ref{figschema}b), and two distinct wells separated by a barrier for $\beta<-1$ (with barrier height controlled by $\Phi_c$, fig. \ref{figschema}c). The flux $\Phi_x$ controls the potential symmetry: for $\Phi_x=0$ the potential is symmetric, otherwise it is tilted (fig. \ref{figschema}c).

In order to study the effect of speed versus adiabaticity on the system, we concentrate our attention on a fast and large modification of the potential shape, from the single-well to the double-well case. This is just an half of the complete manipulation of the qubit state presented in ref. \cite{fastosc1}-\cite{fastosc2}. Initially the system is maintained in the singlewell case for a rest time $t_w$, waiting for the complete relaxation to the ground state. Then it is moved rapidly to the double-well case, where an high barrier separates the two minima, and this is obtained by changing the control flux $\Phi_c$ with a characteristic sweeping rate $\chi=\frac{d \Phi_c}{dt}$. Finally it is done a read out of the SQUID flux state, corresponding to observe which of the two minima is occupied at the end. This is performed by an inductively coupled readout SQUID used as a magnetometer by means of measurements of the switching current \cite{switching}. The sequence is repeated many times in order to estimate the occupation probability $P$ of the final flux (for example, the probability to obtain a final right flux state). The complete operation is repeated for different unbalancing fluxes $\Phi_x$. For slow (adiabatic) modifications we expect that the system remains always in its ground state: the left flux state when the left minima is the lower one (for $\Phi_x<0$), and the right flux state in the opposite case (for $\Phi_x>0$), with a sweet transition between these opposite cases around the symmetry point ($\Phi_x\approx 0$). In this case the probability $P$ as a function of the unbalancing flux $\Phi_x$ presents a sigmoidal shape. By increasing the sweeping rate $\chi$ we expect an excitation of upper levels due to non adiabatic transitions, with a possible emerging of effects related to this population.

\section{Experimental setup and results}
\begin{figure}
\begin{center}
\includegraphics[width=8cm]{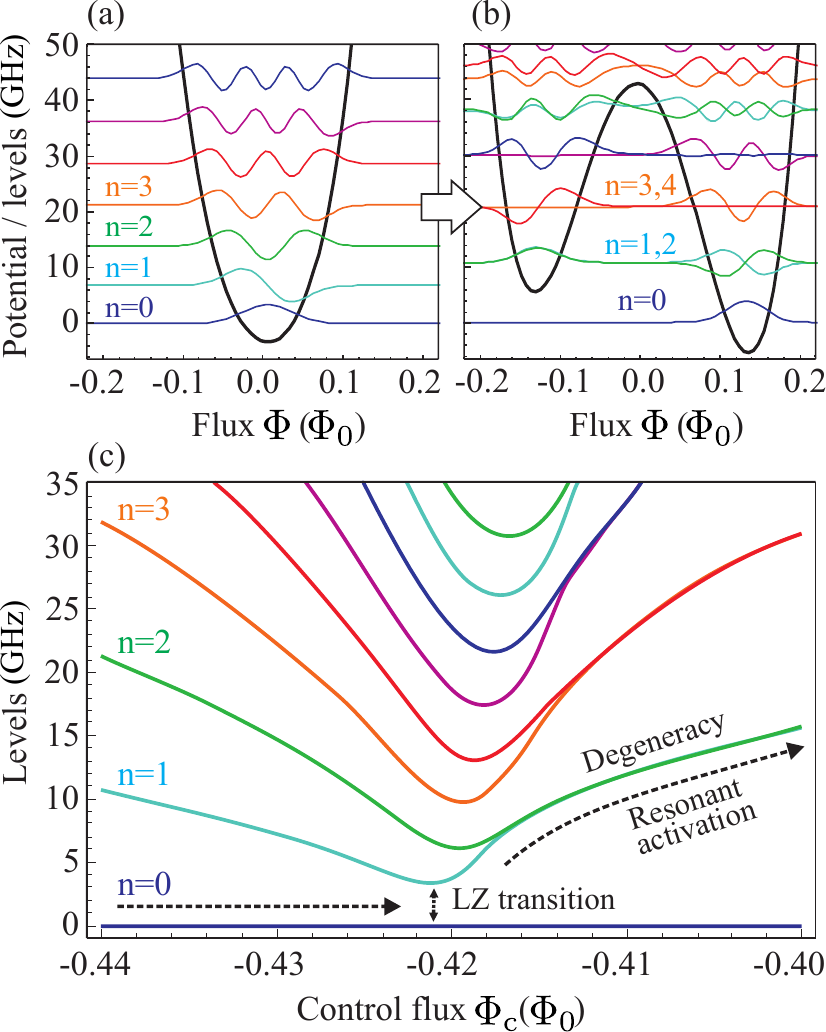}
\caption{(a) Energy potential of the system and relative eigenstates in the single-well case (for $\Phi_c = -429 $ m$\Phi_0$). (b) Energy potential of the system and relative eigenstates in the double-well case (for $\Phi_c = -412 $ m$\Phi_0$), with a slight asymmetry ensuring the degeneracy (for $\Phi_x = 0.543 $ m$ \Phi_0$). (c) Variation of the energy levels positions for different fluxes $\Phi_c$ (for $\Phi_x = 0.543 m \Phi_0$). Note that for convenience all energies are expressed as frequencies in GHz, and are shifted by subtracting the ground state energy.} 
\label{figlivelli}
\end{center}
\end{figure}

We performed the measurements on devices realized by standard trilayer Nb/AlOx/Nb technology, with nominal parameters $L=85pH$, $l=7pH$, $I_0=8\mu A$ and $c=0.3pF$, in a dilution refrigerator with base temperature $T=30mK$ arranged for ultra low noise qubit measurements (mu-metal, superconducting and normal metal shields, thermocoax and L-C-L filters on dc lines, different attenuator stages on the signal line). A preliminary study of the switching current in the readout dc SQUID gives an escape temperature of about 250 mK, compatible with the crossover temperature within the experimental errors. This indicates the absence of an excess temperature due to noise \cite{switching}. The probability $P$ is evaluated by repeating the preparation - modification - readout cycles for 1000 times at a rate of $10 kHz$.
The initial preparation is obtained by waiting for a time $t_w=200 ns$ in the single-well condition (fig. \ref{figlivelli}a), for $\Phi_c\approx -480 m\Phi_0$. In this condition the system is well approximated by an harmonic oscillator with characteristic frequency $\approx 19 GHz$, corresponding to a level spacing of about $0.91 K$, very high with respect to the thermal bath temperature, so that we expect a negligible thermal excitation.

\begin{figure}
\begin{center}
\includegraphics[width=7cm]{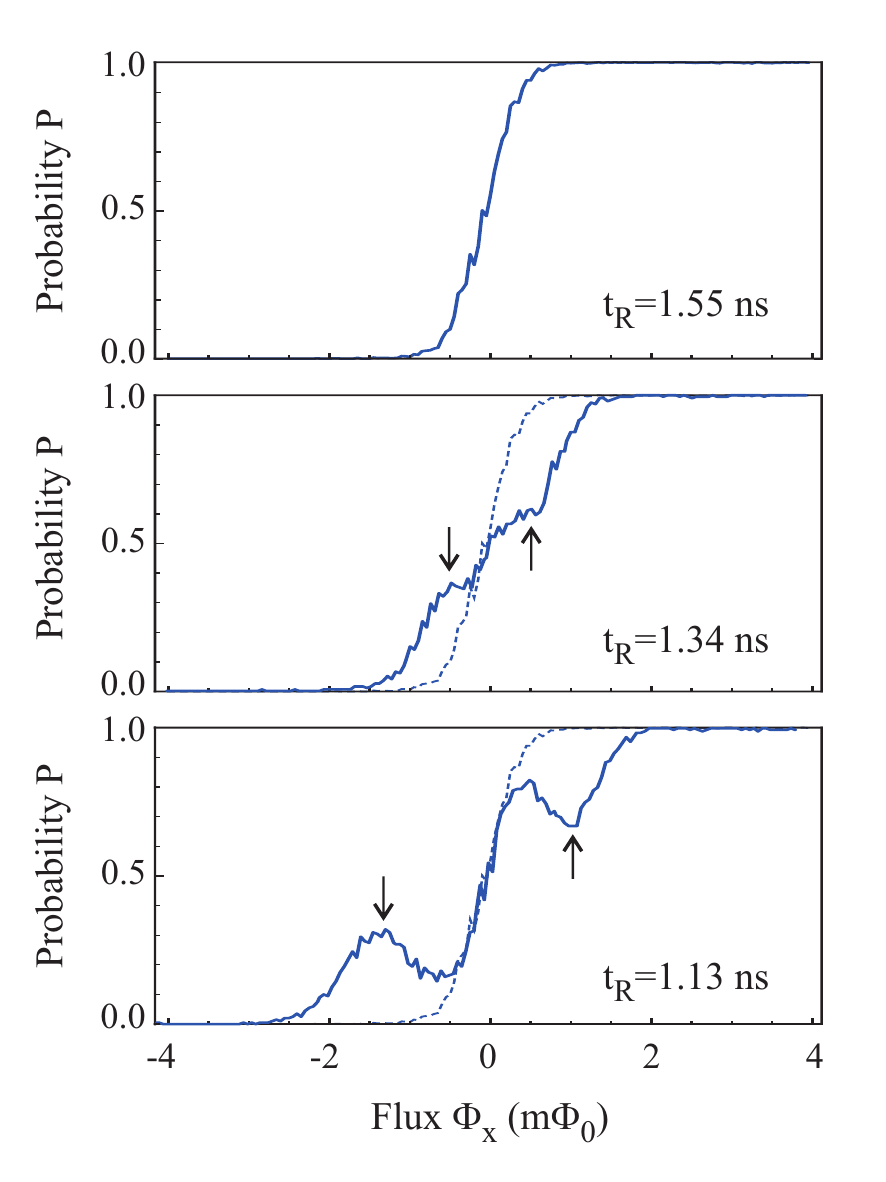}
\caption{Experimental results for the probability $P$ to measure a right flux state for different unbalancings $\Phi_x$ at three different rise times $t_R$. Resonance peaks are visible for faster transitions (arrows in second and third plots)} \label{figmisure2D}
\end{center}
\end{figure}

\begin{figure}
\begin{center}
\includegraphics[width=8cm]{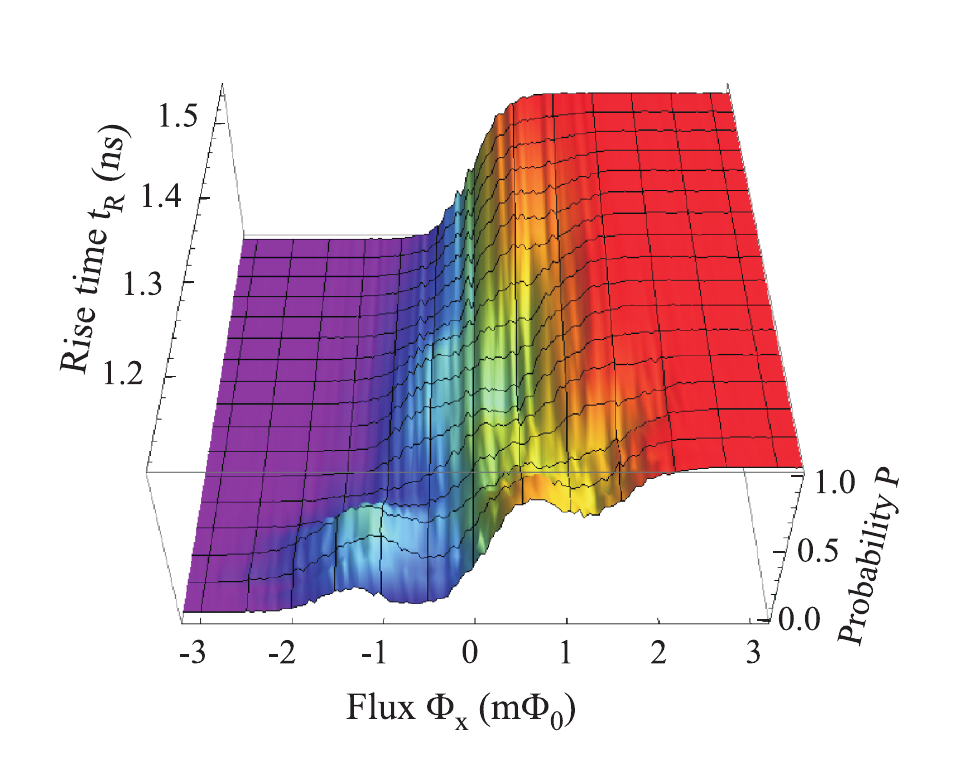}
\caption{Experimental results for the probability $P$ to measure a right flux state for different unbalancings $\Phi_x$ at different rise times $t_R$, plotted as a 3D surface.} \label{figmisure3D}
\end{center}
\end{figure}

The potential shape modification is driven by a fast pulse generator, presenting signals with a typical rise time $t_R=0.8ns$ that can be changed by using an home made tunable L-C-L filter. The modified pulse is fully characterized thanks to a fast oscilloscope, in particular it is possible to check the pulse shape and the actual rise time. The fast signal is transmitted to the device thanks to a $50 \Omega$ matched coaxial cable interrupted by three 20dB attenuators placed at $300 K$, $1 K$ and $30 mK$ stages. More details on the setup can be found in ref. \cite{fastosc1}. We tested the entire line at room temperature and in the absence of the chip, while the present setup does not allow to test the entire system (line plus chip) at low temperature. For this reason we can expect a large error in the determination of the real signal shape and rise time at the device level.
The applied signal modifies the potential from the single-well condition, at $\Phi_c \approx -480 m \Phi_0$, to the double-well case, at $\Phi_c \approx -360 m \Phi_0$, passing through the critical condition $\Phi_c \approx -422 m \Phi_0$ where there is the transition between the single-well and double-well conditions (fig. \ref{figlivelli}).
At the end of each cycle the final flux state is measured by the coupled readout dc SQUID. This is done by applying a current ramp to the SQUID and recording the switching current, which is directly related to the qubit flux.
The sequence is repeated for different unbalancing fluxes $\Phi_x$, ranging from $- 4  m \Phi_0$ to $+ 4  m \Phi_0$, obtaining the probability curves plotted in figure \ref{figmisure2D}. These curves are obtained for three different rise times,  $1.55 ns$, $1.34 ns$ and $1.13 ns$.
In the top plot we observe the sigmoidal function expected for a slow rate. In the middle and lower plots two distinct order of peaks appear, respectively at about $\pm 0.55 m\Phi_0$ and $\pm 1.1 m\Phi_0$. The measurement can be repeated for different rise times obtaining the 3-dimensional curve showed in fig. \ref{figmisure3D}.

From fig. \ref{figmisure3D} we can note some enlightening characteristics.
First of all the position of peaks (in $\Phi_x$) corresponds to the conditions for which different levels in the two wells are aligned (degenerate) (fig. \ref{figlivelli}b). This strongly suggest that the presence of peaks is a manifestation of resonant tunneling between wells.
Secondly we note that the appearing of peaks requires rise times below a particular critical value, namely it is necessary to have an high enough sweep-rate in order to observe peaks.
Thirdly, there is an alternation of  peaks' orders: when the second order of peaks appears the first order disappears.
In fig. \ref{figlivelli}c it is plotted the modification of the first nine energy levels in the passage from the single-well to the double-well condition (in the degenerate case for $\Phi_x = 0.55 m \Phi_0$). This figure can help us in a qualitative explanation of the observed peaks. In the single-well condition (on the left) it appears reasonable to suppose that only the ground state is populated. Close to the critical point $\Phi_c \approx -0.42 \Phi_0$, where the barrier appears to separate two distinct wells, Landau-Zener transitions populate the upper levels, with an efficiency depending on the sweep-rate $\frac{d \Phi_c}{dt}$. These excited states can cross the barrier thanks to a resonant tunneling when the alignment condition is meet.   The experimental results thus suggest that these two effects combine together and produce the observed peaks, due to an excess of population in the upper well when the resonant and the non-adiabatic conditions are both fulfilled. The region where this effect is active is small, of the order of $1/10$ of the entire span of the flux $\Phi_c$ ($120 m \Phi_0$), and this region is crossed in a similar fraction of the entire rise time duration, about $0.1 ns$. 
We stress again that the combination of Landau-Zener and resonant tunneling can explain the first two observations (position of peaks and appearing of them only below a critical rise time), but not the third one, that is the alternation of peaks' orders.
\section{Simulations and discussions}
In order to gain information about possible physical mechanisms and/or properties of the system which may be responsible for the appearance of the alternation of peaks' order as in fig. \ref{figmisure3D}, in what follows we develop a more quantitative analysis exploiting a simple quantum model useful to describe the system under scrutiny. To do this we start considering the Hamiltonian model relative to the potential (\ref{H}), rewritten in a more convenient way:

\begin{equation}\label{MainHamilt}
H(t)=-\frac{1}{2m}\frac{\partial^{2}}{\partial \varphi^{2}} + m
\Omega^{2}\frac{(\varphi-\varphi_{x})^{2}}{2}+ m
\Omega^{2}\beta(t) \cos(\varphi)
\end{equation}

where $\Omega={1/\sqrt{LC}}$, $\varphi=\Phi/{\Phi_B}$ and  $\varphi_{x}=\Phi_x/\Phi_B$.  Moreover $\beta(t)=-I_0(\Phi_c(t))\frac{\Phi_B}{m\Omega^2}$. The Hamiltonian (\ref{MainHamilt}) is well suited to describe the time evolution of a particle in one dimensional time-dependent potential. In particular, appropriately choosing the function $\beta(t)$, it is possible to vary the potential shape going, in an interval of time $\tilde{t}$ from a single to a double-well, thus reproducing the initial and final conditions of the experiment before discussed. The problem therefore consists in finding, as function of both the unbalancing parameter $\varphi_x$ as well as of the rise time $t_R$, the probability $P$ that, at the end of the process, the particle is found in the right well. Let's observe that knowing the state of the system  $\ket{\Psi(\tilde{t})}$ at the time instant $\tilde{t}$, this probability can be simply evaluated as 

\begin{equation}\label{Pl}
P=\int_{\textrm{right well}} \Psi^*(\tilde{t})\Psi(\tilde{t})d \varphi
\end{equation}
Let's indicate by $\ket{\psi_n(t)}$ a set of instantaneous eigenfunctions of the Hamiltonian (\ref{MainHamilt}):
\begin{equation}\label{eigenstates}
H(t)\ket{\psi_n(t)}=E_n(t)\ket{\psi_n(t)}
\end{equation}
Exploiting these states, we can write
\begin{equation}\label{Psi_t}
\ket{\Psi(t)}=\sum_{n=0}^{\infty}\exp[i\int_0^tE_n(\tau)d\tau] \cdot s_n(t) \cdot \ket{\psi_{n}(t)}.
\end{equation}
where the function $s_n(t)$ are solutions of the following set of integro-differential equations:
\begin{equation} \label{main diff eq}
\dot{s}_n(t)=-\sum_{k=0}^\infty M_{nk}(t)\exp[i\int_0^t(E_n(\tau)-E_k(\tau))d\tau]s_k(t).
\end{equation}
with  $M_{nk}(t)=\bra{\psi_{n}(t)}\dot{\psi}_{k}(t)\rangle$. Starting from eqs. (\ref{Pl}) and (\ref{Psi_t}), the probability $P$ can be then written as

\begin{eqnarray}\label{PlEigenF}
P&&=\sum_{n=0}^{\infty}|s_n(\tilde{t})|^2\int_{\textrm{right well}}\psi_n^*(\tilde{t})\psi_n(\tilde{t}) d\varphi \\ \nonumber
&&\equiv \sum_{n=0}^{\infty}|s_n(\tilde{t})|^2 L_n
\end{eqnarray}
with
\begin{equation}\label{Ln}
L_n=\int_{\textrm{right well}}\psi_n^*(\tilde{t})\psi_n(\tilde{t}) d\varphi 
\end{equation}

 An exact analytical resolution of the coupled integro-differential equations (\ref {main diff eq}) is not easy. Thus we proceed further by performing  numerical simulations of the dynamical behavior of the system carefully  taking into account both the non adiabadicity in the system dynamics and the possible  emergence of resonant tunneling processes.  As first step, considering $\beta(t)$ as a parameter, we numerically diagonalize the Hamiltonian (\ref{MainHamilt}) at a generic time instant $t$, finding its instantaneous eigenvectors $\ket{\psi_n(t)}$ and the correspondent eigenvalues $E_n(t)$. 

Considering values of $t_R$ of interest in the context of this paper, we have evaluated the quantity $L_n$, defined in eq. (\ref{Ln}), in correspondence to different values of $n$ verifyng,  as expected, that, at least for not too large $n$, $L_n$ is almost equal to one or negligible witnessing that the first eigenstates are practically localized for $\Phi_x\neq 0$. In our simulation however we use the numerical value of $L_n$ instead of 0 or 1.

\begin{figure}
\begin{center}
\includegraphics[width=8cm]{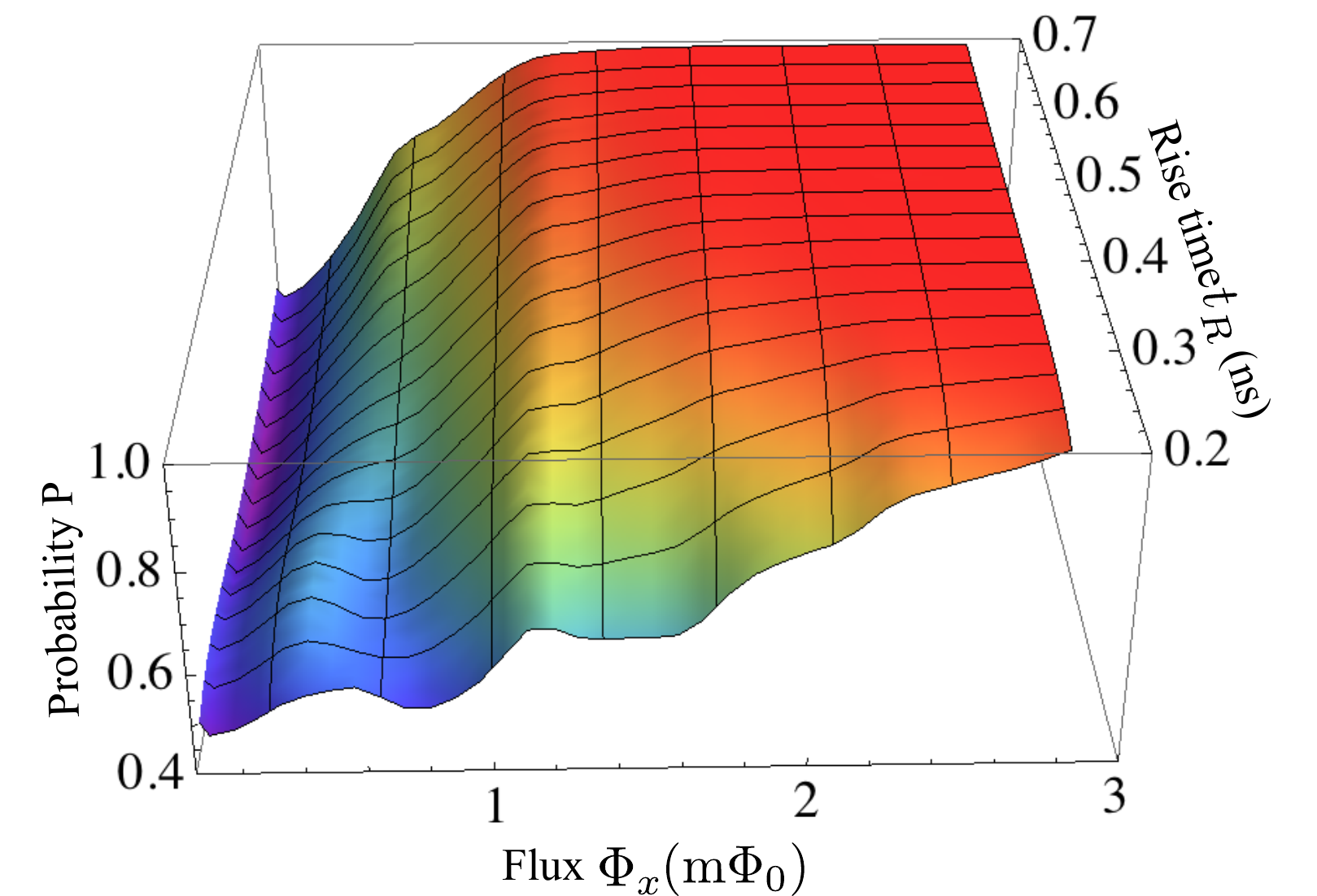}
\caption{Probability $P$ to find the system in the right well against the unbalancing $\Phi_x$ and rise time $t_R$, supposing that at $t=0$ it is in its ground state.} \label{ground}
\end{center}
\end{figure}

To evaluate the probability $P$  we thus need to calculate the populations $|s_n (\tilde{t} )|^2$ by numerically solving the system (\ref{main diff eq}) explicitly giving the way in which the potential shape modifies itself going from the initial condition to the final one during the time $\tilde{t}$. In other words we now have to choose the function $\beta(t)$ appearing in eq. (\ref{MainHamilt}).  We wish to underline that this is a very delicate point. It is undoubted indeed that the dynamics of the system will be deeply affected from the way of varying the potential shape. Thus we expect to find different results in correspondence to different choice of $\beta(t)$.  

A sigmoidal function allows a simple and reasonable description, at least from a qualitative point of view because of experimental uncertainties on the exact $\beta(t)$  shape as discussed in section III.
Thus we fix it as  $\beta(t)=\beta(0)(1-\zeta(t))+\beta(\tilde{t})\zeta(t)$ with $\zeta(t)=\frac{Erf((2t/\tilde{t}-1)w-s)-Erf(-w-s)}{Erf(w-s)- Erf(-w-s)}$,  $Erf(x)=\frac{2}{\sqrt{\pi}}\int_0^x e^{-t^2}dt$, choosing in particular $w=2$ and $s=0.3$. It is the case to stress that varying these two parameters implies, as a consequence, a changing in the rise time $t_R$.
In order to investigate on the appearance of peaks in fig. \ref{figmisure3D}, we have calculated the probability of finding the particle in the right well for different  values of $\Phi_x=\varphi_x\Phi_B$ supposing, as it appears physically reasonable, that at $t=0$ the system is in its ground state. In particular we evaluated such a probability considering a range of  $\Phi_x$ in correspondence of which peaks appear, as suggested by the experimental data.  Considering values of $t_R$ like in figure \ref{figmisure3D}, our simulation does not evidence the existence of significant peaks. However, taking into account the fact that the peaks in the probability $P$ arise reducing the rise time and in view of the experimental uncertainties before discussed, we have simulated the behavior of the system exploring smaller $t_R$. 
The results obtained are reported in figure (\ref{ground}) where the probability $P$ is plotted as function of both $\Phi_x$ and $t_R$.

\begin{figure}
\begin{center}
\includegraphics[width=8.5cm]{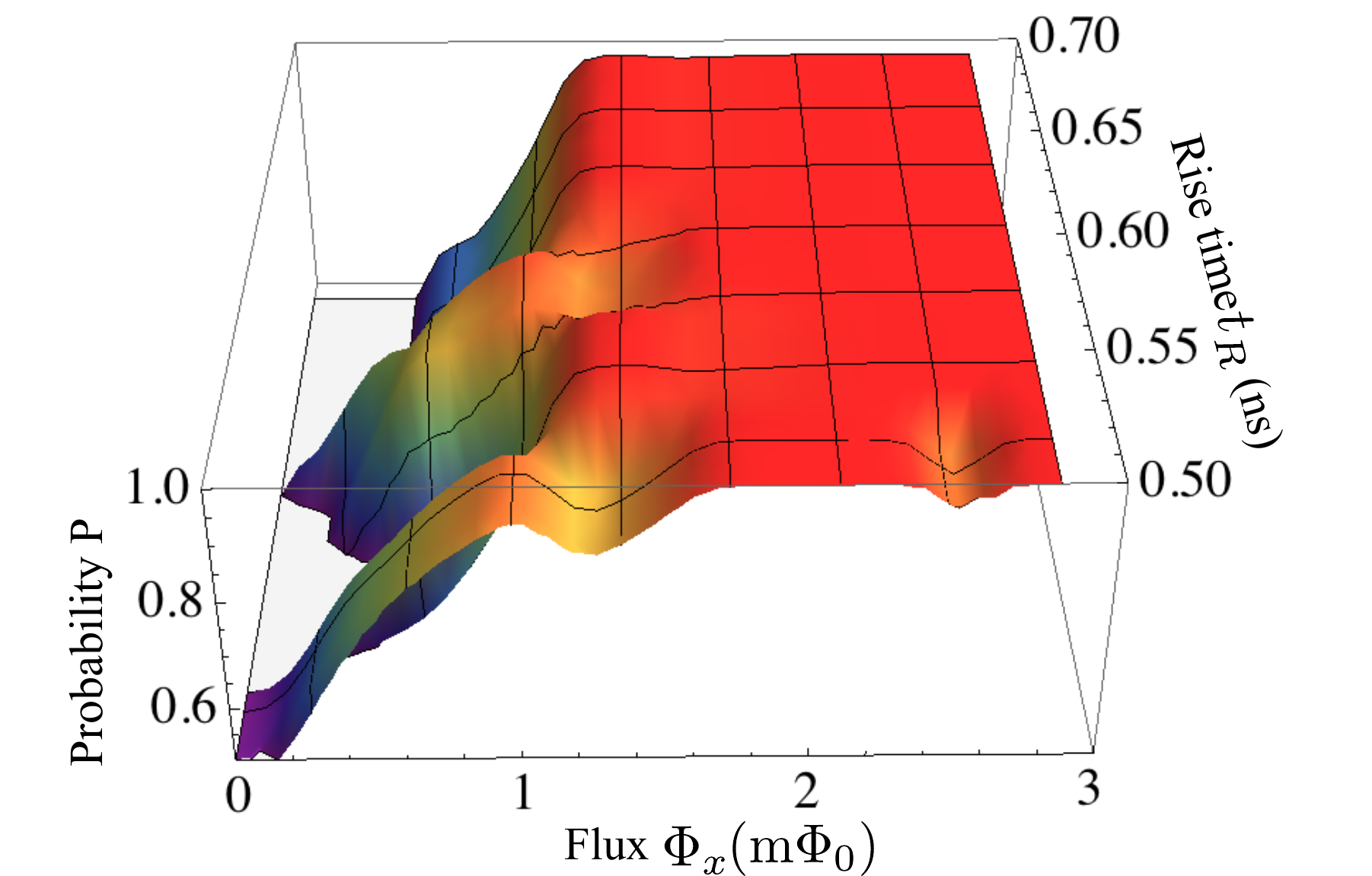}
\caption{Probability $P$ to find the system in the right well against the unbalancing $\Phi_x$ and rise time $t_R$, supposing that at $t=0$ it is in the linear superposition $\ket{\Psi(0)}=x\ket{\psi_0(0)}+\e^{i\theta}\sqrt{(1-x^2)}\ket{\psi_1(0)}$ of its ground  and first excited states in correspondence to $x=0.95$ and $\theta=0$.} \label{combinazione1}
\end{center}
\end{figure}

As expected, peaks of resonance appear in correspondence to different values of  $\Phi_x$. However the position of such peaks with respect to the rise time $t_R$ does not reflect, not only quantitatively but also qualitatively, the experimental observations. In other words even if a dependence of $P$ on $t_R$  is evident, the function $P(t_R)$   is very different from the experimental one.  In particular the probability $P$ obtained by simulation (see fig. \ref{ground}) is not characterized by the alternation of  peaks' order as in figure \ref{figmisure3D}.  
Such result moreover does not seem to be imputable to the particular choice of the $\beta(t)$ we made. We have indeed verified that this is the case by choosing a linear function and obtaining the same qualitative behavior of that shown in fig. \ref{ground}.
If it is true that the manner in which we modify the potential shape deeply affects the dynamics of the system and thus the probability $P$ that the system is found in the right well at the end of the potential modification, another key ingredients to be considered is surely the state of the system at $t=0$. Taking into account the fact that the temperature at which the experiment is performed is  $\sim 30 mK$, it is  reasonable to suppose that at $t=0$ there is a small, but not zero (of the order of few percent), probability that the system is in its first excited state.  Thus could be reasonable to assume that the preparation step leaves the system in a mixture $\rho=x^2\ket{\psi_0(0)}\bra{\psi_0(0)}+(1-x^2)\ket{\psi_1(0)}\bra{\psi_1(0)}$ of the ground and the first excited state. Performing simulation starting from  this mixture instead of the ground state, we have verified that, considering values of $x$ compatible with $T\simeq30mK$, we do not get significant differences with respect to the results displayed in fig. \ref{ground}. We have also checked that increasing the relative weight \textit{x} of the first excited state in the initial mixture worsens the accordance between theoretical predictions and experimental results.
The theoretical prediction instead drastically changes  if we suppose that quantum coherences are present in the initial state of the system. Such an assumption can be justified by considering the fact that the waiting interval of time $t_w$  was not long enough to allow the complete destruction of the coherences between the ground and the first excited state of the double SQUID. If this is the case it is reasonable to assume that at $t=0$ the system is in a quantum superposition $\ket{\Psi(0)}=x\ket{\psi_0(0)}+\e^{i\theta}\sqrt{(1-x^2)}\ket{\psi_1(0)}$  of the first two low-lying states, instead of a mixture of the two states as before supposed. Starting from this initial state the probability $P$ shows a dependence on both $\Phi_x$  and  $t_R$ as displayed in fig. (\ref{combinazione1}) where we have considered a smaller range of $t_R$ to better appreciate the behavior of $P$.
As expected, also in view of experimental uncertainties on the $\beta(t)$ function as well as on the parameters defining the system, figure (\ref{combinazione1}) does not exactly match the experimental results presented before,  even if  the qualitative behavior of $P$ seems to be well reproduced. More in detail the most important aspect of the results shown in fig. (\ref{combinazione1}) consists in the fact that, as experimentally observed, there is an alternation of the peaks' order determined by both the asymmetry in the potential governed by the value of the unbalancing parameter $\Phi_x$, and on the rise time $t_R$ required to go from a single to a double-well. 
The theoretical analysis developed in this paper has the merit to disclose the role played by the persistence of quantum coherences in the initial state of the double SQUID.  We wish to stress indeed that starting from an initial state as the ground state of the qubit or a mixture of the same ground state and the first excited one, even if leading to the appearance of peaks, is completely unable to predict the alternance of minima and maxima as requested by the experimental results. Thus our  assumption, that is the persistence of quantum coherences, leads to predictions in good qualitative agreement with the experimental results. The intriguing point is that all the alternative seemingly more reasonable assumptions concerning the initial state of the SQUID predict a behavior not compatible with some aspect of the observed one.

Before concluding we wish to spend some words about possible decoherence effects in the dynamical behavior of the system. It is important to underline that, as we have previously discussed, the temporal interval where the physical mechanisms at the basis of the observed effects, are active is a small fraction (of the order of 0.1 ns) of the total duration of the experiment (about 1 ns).  As first approximation it is thus reasonable to neglect decoherence effects in the system dynamics, which present time scales of the order of nanoseconds.  Anyway it is the case to underline that in this context the source of noise is the $\frac{1}{f}$ noise \cite{Chiarello13} that generally speaking acts modifying  the effective control parameter as for example $\Phi_x$ \cite{decoherence1}-\cite{decoherence3}. We thus expect that  the effects of such a noise on the results reported in figure \ref{combinazione1} would consist  at most in a broadening of the observed peaks. 

Summarizing, we have experimentally verified that a fast modification of the potential shape of a double SQUID gives rise to the activation of quantum resonance phenomena manifesting themselves as peaks in the probability of measuring a right flux state of the SQUID at the end of the non adiabatic transition. The theoretical analysis we have performed seems moreover to lead to  the conclusion that the characteristic behavior of such a probability is also a direct consequence of the presence of quantum coherences in the initial state of the system. 
 This fact suggests to perform other experiments on the system under scrutiny aimed at revealing quantum interference effects in the behavior of some no-diagonal physical observables.

\begin{acknowledgments}
The authors  thank Dr. Evgeniy Safonov  for stimulating and deep
discussions on the subject of this paper. Financial support from the Italian Project PRIN
2008C3JE43\_003 is acknowledged.
\end{acknowledgments}

\end{document}